\documentclass[%
nofootinbib,reprint,
amsmath,amssymb,
aps,
]{revtex4-2}
\usepackage{easyReview}
\usepackage{graphicx}
\usepackage{dcolumn}
\usepackage{bm}
\usepackage{color}
\usepackage{amsmath}
\usepackage{epsfig,graphics}
\usepackage{etoolbox}
\usepackage{hyperref}
\usepackage{gensymb}
\usepackage{slashed}

\hypersetup{colorlinks}
\usepackage{physics}

\definecolor{red}{rgb}{0.9, 0,0}
\definecolor{cerulean}{rgb}{0., 0.62,0.9}
\definecolor{navy}{rgb}{0.05, 0.05,0.8}
\definecolor{orange}{rgb}{1., 0.65,0.0}
\definecolor{magenta}{rgb}{1.0, 0.08, 0.58}

\newcommand{\miss}{\text{miss}}

\newcommand{\gammainv}{\gamma_{\rm inv}}
\newcommand{\GeV}{\text{ GeV}}

\newcommand{\thetaestar}{\theta_e^*}

\newcommand{\pLmiss}{  p_{ \text{miss}}^L  }

\newcommand{\BelleII}{Belle~II }

\makeatletter
\patchcmd\@footnotemark{{link}}{{footnote}}{}{\fail}
\makeatother
\ActivateGenericHook{hyp/link/footnote}
\AddToHook{hyp/link/footnote}{\hypersetup{linkcolor=cyan}}
\begin{document}

\title{Fusing photons into nothing,\\ a new search for invisible ALPs and Dark Matter at Belle II}

\author{Francesca Acanfora$^\dagger$$*$} 
\author{Roberto Franceschini$^\dagger$} 
\author{Alessio Mastroddi$^\dagger$}
\author{Diego Redigolo$^{\ddagger}$}
\affiliation{
    $^\dagger $ Universit\`a degli Studi and INFN Roma Tre, Via della Vasca Navale 84, I-00146, Rome, Italy\\
    $^*$ Galileo Galilei Institute for Theoretical Physics, Largo Enrico Fermi 2, I-50125 Firenze, Italy \\
    $^\ddagger$ INFN, Sezione di Firenze Via G. Sansone 1, 50019 Sesto Fiorentino, Italy, \\
}%


\begin{abstract}
    We consider an axion-like particle coupled to the Standard Model photons and decaying invisibly at Belle~II. We propose a new search in the $e^+e^-+\text{invisible}$ channel that we compare against the standard $\gamma+\text{invisible}$ channel.  We find that the $e^+e^-+\text{invisible}$ channel has the potential to ameliorate the reach for the whole  ALP mass range. This search leverages dedicated kinematic variables which significantly suppress the Standard Model background. We explore the implications of our expected reach for Dark Matter freeze-out through  ALP-mediated annihilations.
\end{abstract}

\maketitle


\section{Introduction}\label{sec:introduction}

Dark Matter (DM) searches at the intensity frontier are like a fishing expedition in the high sea at a depth never explored before. Going to high intensity opens up the possibility to test directly extremely feebly interactions of DM  with the Standard Model which would be impossible to probe otherwise. These interactions can be responsible to produce light DM (in the $\text{MeV} -\text{GeV}$ range) in the early Universe through thermal freeze-out~\cite{Krnjaic:2022ozp}.

Collider searches support direct detection experiments and indirect detection observations in the joint effort of testing the allowed parameter space of thermal DM freeze-out. This complementarity is particularly important for specific kinematic configurations suppressing the DM elastic scattering with the target materials or the DM annihilation in the galactic environment~\cite{Essig:2022dfa,Leane:2020liq}.

\begin{figure}[t!]
    \centering
    \includegraphics[width=0.99 \columnwidth]{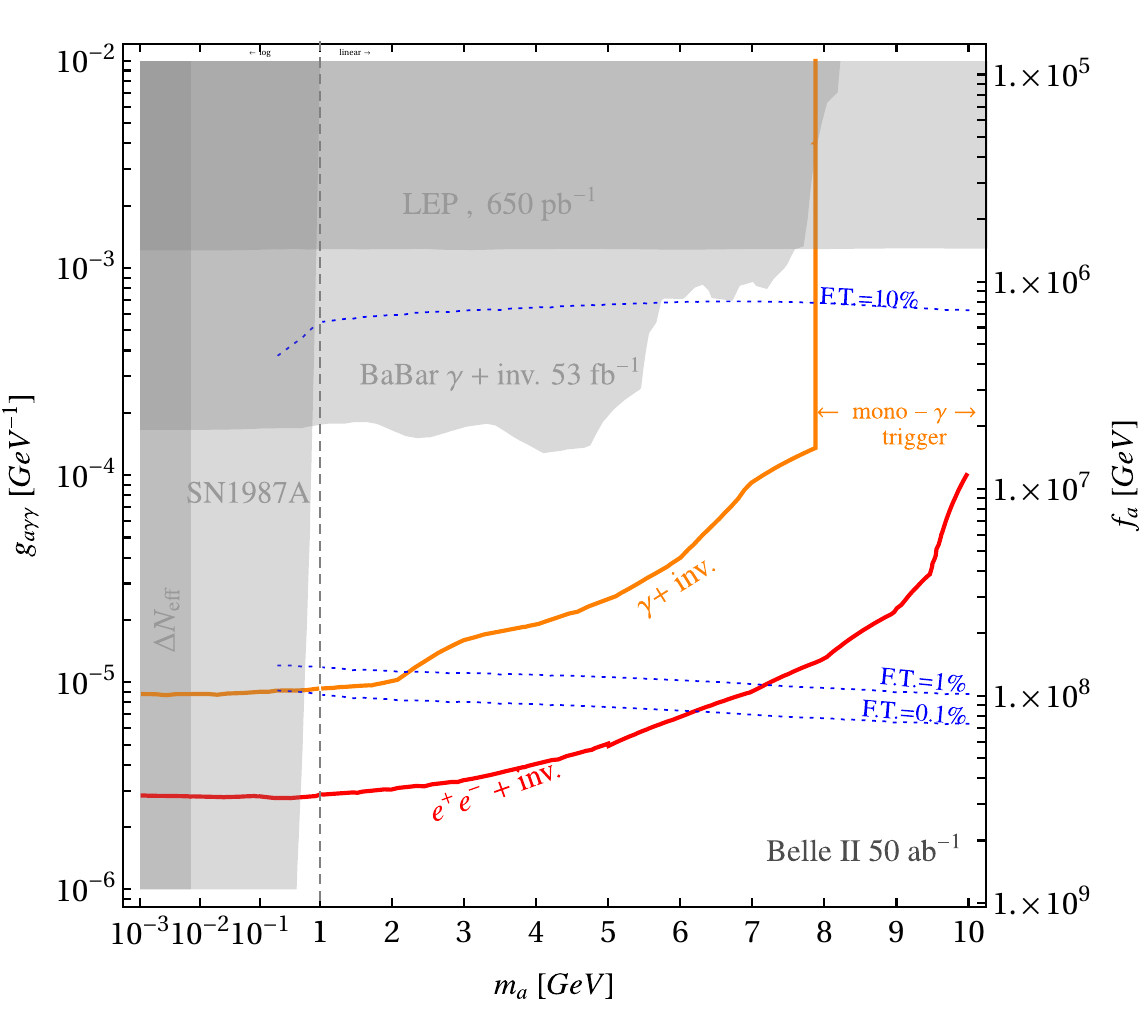}
    \caption{Expected sensitivity of \BelleII at 95\% C.L. to the ALP coupling to photons $g_{a\gamma\gamma}$ as defined in Eq.~\eqref{eq:ALPlagrangian}. The branching ratio of the ALP into invisible states is taken to be one as motivated by Eq.~\eqref{eq:power counting}. The {\bf orange} line is the expected reach of the $\gamma+\text{invisible}$ channel derived in Ref.~\cite{Dolan:2017osp}. The {\bf red} line shows the reach of the $e^+e^-+\text{invisible}$ channel discussed here. The {\bf gray shaded} region shows existing constraints from LEP and Babar $\gamma+\text{invisible}$ searches~\cite{BaBar:2017tiz,Fox:2011fx,DELPHI:2008uka} and from $\Delta N_{\text{eff}}$ constraints from CMB~\cite{Boehm:2013jpa}. We also show the expected constraint from SN cooling estimated in Ref.~\cite{Dolan:2017osp}. The {\bf dotted blue} lines show the freeze-out prediction for resonant DM annihilation with fine tuning (F.T.) $10\%,1\%$ and $0.1\%$ as discussed below Eq.~\eqref{eq:resonant}.}\label{fig:money} 
\end{figure}

In this paper we revisit the sensitivity of the \BelleII experiment to DM communicating with the SM through an axion-like particle coupled only to photons. This simple dark sector scenario was considered in Ref.~\cite{Dolan:2017osp}, where the sensitivity of \BelleII was derived focusing on the standard mono-photon final state accompanied by missing energy. The same experimental strategy was implemented before at BaBar~\cite{BaBar:2008aby,BaBar:2010eww,BaBar:2017tiz} and it is under implementation of the \BelleII collaboration~\cite{Belle-II:2018jsg} and expected to give results soon~\cite{Zanitalk}. 

We develop an alternative strategy based on the 
$$e^+e^- \to e^+e^-+\text{invisible}$$
channel, leveraging a more detailed knowledge of the signal kinematics (given the two visible particles) at the price of the reduced production cross section. In Fig.~\ref{fig:money} we show our main result, demonstrating how this search strategy can provide a new powerful and independent probe of this type of new physics. 

A key observation behind our strategy is that a system of multiple invisible particles, as it is the typical for SM backgrounds, is not likely to have \emph{large} missing energy, and at the same time a \emph{small} invariant mass and a \emph{small} longitudinal missing momentum with respect to the missing energy of the system. On the contrary, this kind of kinematics is peculiar of a single invisible body of small mass such as an ALP decaying invisibly.

The paper is organized as follows. In Sec.~\ref{sec:setup} we specify the theoretical setup, discuss its parameter space and the ancillary constraints from cosmology, astrophysics, direct and indirect detection. In Sec.~\ref{signal} we compare the $\gamma+\text{invisible}$ and $e^+e^-+\text{invisible}$ final states at \BelleII. We discuss the signal cross sections and the signal kinematics. Additional information about the signal is given in App.~\ref{app:signal}. In Sec.~\ref{qed-zero} we characterize the SM backgrounds with further details given in App.~\ref{app:bkd}. In Sec.~\ref{sec:event-selection} we specify the event selection that lead to the expected sensitivity in Fig.~\ref{fig:money}. In Sec.~\ref{discussion} we discuss our result, the necessary further steps to complete our study and the possible future directions.  

\section{Setup}\label{sec:setup}

For concreteness and to characterize our experimental reach we specify the DM model with an ALP mediator already considered in Ref.~\cite{Dolan:2017osp}. The Lagrangian reads\footnote{Notice that $g_{a\chi\chi}$ in Eq.~\eqref{eq:ALPlagrangian} is related to one defined in Ref.~\cite{Dolan:2017osp} as ${\cal L}\supset\tilde{g}_{a\chi\chi}\bar{\chi}\gamma^{\mu}\gamma_5\chi\partial_\mu a$ by a simple redefinition $g_{a\chi\chi}=2 \tilde{g}_{a\chi\chi}$ upon integration by parts and DM equation of motion.}  
\begin{widetext}
    \begin{equation}
        {\cal L}= \dfrac{1}{2} (\partial_\mu a)^2-\dfrac{m_a^2 }{2} a^2 - \frac{g_{a\gamma\gamma}}{4} a F_{\mu\nu}\tilde{F}^{\mu\nu}+\frac{i}{2}\bar{\chi}\gamma^\mu\partial_\mu\chi+\frac{M_\chi}{2}\bar{\chi}\chi+\frac{g_{a\chi\chi}}{2} M_\chi a \bar{\chi}\gamma_5\chi \ ,\label{eq:ALPlagrangian}
    \end{equation}
\end{widetext}
where $\chi$ is a Majorana fermion.

If the ALP is the pseudo-Nambu-Golstone boson (pNGB) of a global $U(1)$ symmetry, we can estimate the size of its coupling to photons and DM in terms of the decay constant $f_a$ which controls the UV cutoff of the theory $\Lambda_{\text{UV}}=g_* f_a$, where $g_*$ is an $\mathcal{O}(1)$ coupling. In this setup, the ALP coupling to photons originates from the ABJ anomaly of the  global $U(1)$ symmetry with respect to QED and 
\begin{equation}
    g_{a\gamma\gamma}\equiv\frac{\alpha_{\text{em}} c_{\gamma\gamma}}{2\pi f_a}\ , \label{eq:power-counting}
\end{equation}
where the anomaly coefficient $c_{\gamma\gamma}=\sum_E Q_E^2$ is controlled by the charge $Q_E$ of chiral fermions of mass $\Lambda_{\text{UV}}$.  

The ALP coupling to DM can be generated through explicit breaking of the ALP shift symmetry. As a consequence, the ALP mass is naturally of the same order of the DM one, motivating a mass hierarchy like the one considered in this paper, where the ALP mass is slightly heavier than the DM one.

In such a theory, the ALP decays invisibly into DM pairs with a branching ratio close to 1 since the decay into a pair of photons is loop-suppressed. The ratio of the partial decay widths can be estimated to be
\begin{equation}
    \frac{\Gamma(a\to\gamma\gamma)}{\Gamma(a\to \chi \chi)}\sim\left(\frac{\alpha_{\text{em}}}{4\pi}\right)^2\frac{1}{r^2\sqrt{1-4 r^2}}   \ ,  \label{eq:power counting}
\end{equation}
with $r\equiv M_\chi/m_a\lesssim 1/2$ and for $c_{\gamma\gamma}\sim\mathcal{O}(1)$ and $g_{a\chi\chi}\sim1/f_a$. Where the overall $1/r^2$ factor comes from the scale dependence of the partial width while $\sqrt{1-4 r^2}$ accounts for the possible phase space suppression of the invisible channel. The suppression of the diphoton width in Eq.~\eqref{eq:power counting} is generically of order $\mathcal{O}(10^{-7}-10^{-6})$ and can only be compensated in scenarios where either the DM mass is very light, i.e. $r\to0$, or the DM mass $M_\chi$ is very close to $m_a/2$, i.e. $r\to1/2$. 

The simple theory in Eq.~\eqref{eq:ALPlagrangian} can be used as a model of DM freeze out through ALP-mediated annihilations into photons as first studied in Ref.~\cite{Dolan:2017osp}. Setting the relic abundance of $\chi$ to match the measured DM abundance today~\cite{Steigman:2012nb} generically requires $f_a$ to be around the GeV with a mild dependence on the DM mass. This region is already excluded by existing collider searches. 

A simple way to push the freeze-out region at weaker coupling is to make the annihilation resonant by tuning the DM mass to be close to the ALP resonance (i.e. $r\to1/2)$. As first discussed in Ref.~\cite{Dolan:2017osp}, defining $x=M_\chi/T$ the thermally averaged annihilation cross section in this limit can be approximated as 
\begin{equation}
    \left\langle\sigma v\right\rangle\approx \frac{\pi}{64}\cdot g_{a\gamma\gamma}^2 \cdot \frac{x}{r^{5}}\frac{K_1(x/r)} {K_2(x)^2}  \ ,\label{eq:resonant}
\end{equation}
which is independent on $g_{a\chi\chi}$ as long as the total width of the ALP can be approximated with its invisible decay width. The prediction of resonant freeze-out are shown in Fig.~\ref{fig:money} for fine-tuned values of $r$ close to the resonance, where $\text{F.T.}\equiv 1/2-r$ can be taken as a measure of the fine-tuning. Assuming instantaneous freeze-out we impose  $\left\langle\sigma v\right\rangle\vert_{x_*}=\sigma_{\text{f.o.}}$ using the results of Ref.~\cite{Steigman:2012nb} for the freeze-out temperature $x_*$ and for $\sigma_{\text{f.o.}}$. Interestingly, the boost to the annihilation cross section given by the resonance saturates for fine-tunings smaller than $10^{-3}$ making it possible to define a \emph{minimal} $g_{a\gamma\gamma}$ compatible with thermal freeze-out  which is shown in Fig.~\ref{fig:money}. Remarkably, the reach of \BelleII based on the $e^+e^-+\text{invisible}$ channel proposed here could be able to probe this coupling for DM masses below 6 GeV. 

Given the average velocity of DM in the Milky Way, which correspond to $x\sim 10^6$, the ALP-mediated annihilation cross section today is suppressed by the asymptotic of the Bessel functions in Eq.~\eqref{eq:resonant}. As a result the present constraints from indirect detection (see e.g. Ref.~\cite{Leane:2018kjk}) cannot test resonant thermal freeze-out. The direct detection reach can be extracted by integrating out the ALP and writing dimension 7 operators coupling the DM to the photon bilinear. The rate of DM scattering onto nuclear and electron targets is heavily suppressed by the high dimension of the operators mediating the DM scattering so that the reach of direct detection is not constraining the parameter space of resonant ALP-mediated freeze-out~\cite{Weiner:2012cb,Frandsen:2012db,Kavanagh:2018xeh}. Colliders are then our best hope to test such a peculiar scenario of DM production.

For low ALP masses, strong constraints come from the measurements of the effective number of relativistic species at BBN and CMB. In our setup the strongest bound can be derived from current Planck measurements~\cite{Planck:2018vyg} by computing the DM entropy transfer to the electron-photon bath after neutrino decoupling~\cite{Boehm:2012gr,Chu:2022xuh}. We show this constraint in Fig.~\ref{fig:money} which robustly rules out DM masses below roughly 10 MeV. 

Stronger constraints could be derived by requiring the ALPs produced in the nascent proto-neutron star (PNS) during the supernova (SN) explosion to not substantially modify the canonical neutrino cooling  mechanism~\cite{Raffelt:1990yz}. This constraint assumes that the DM mean-free-path is longer than the size of the PNS $(\sim10 \text{ km})$ which is always the case in the parameter space of interest. Setting a rigorous bound for heavy ALP masses (beyond the CMB bound) depends very much on the parameters controlling the SN thermodynamics and goes beyond the scope of this work. In Fig.~\ref{fig:money} we show the approximate line derived in ~\cite{Dolan:2017osp} as an indication of the possible constraining power of this observable.

\begin{figure*}
    \centering
    \includegraphics[scale=.6]{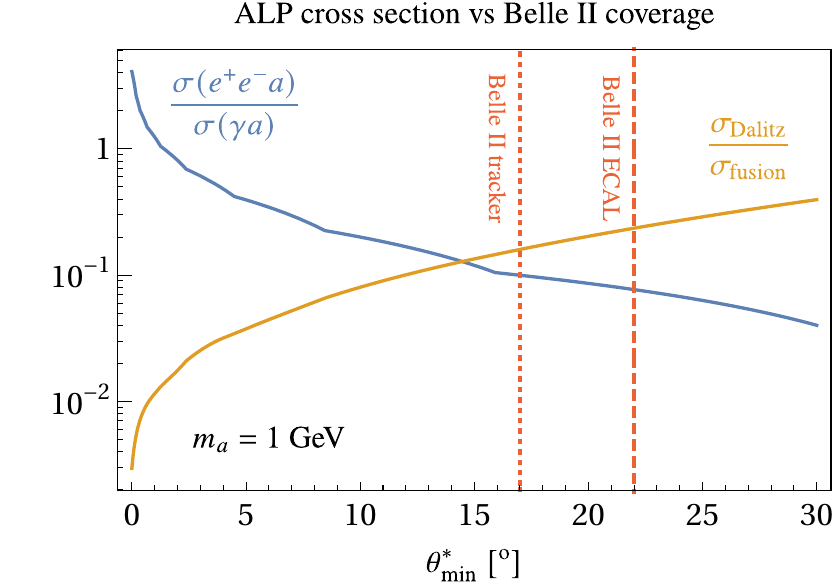}\hfill
    \includegraphics[scale=.6]{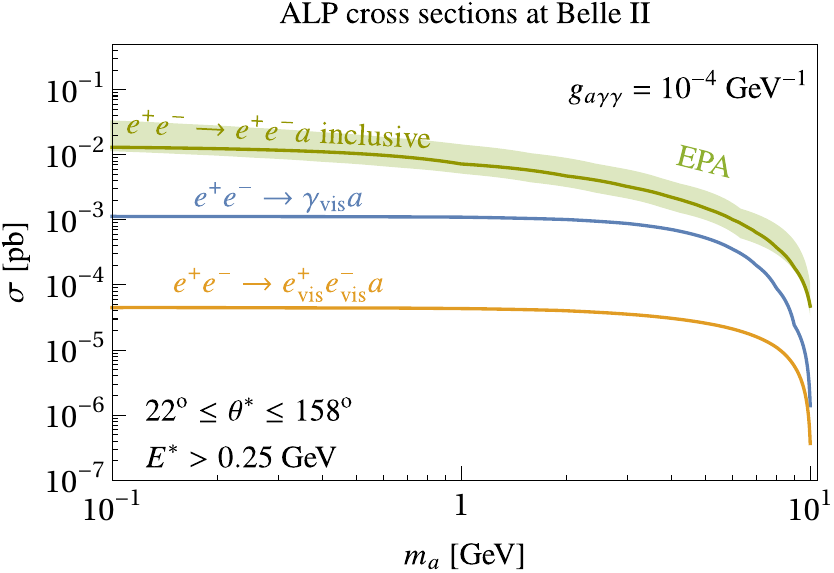}
    \caption{Characterization of the different production channels for an ALP coupled to SM photons as defined in Eq.~\eqref{eq:ALPlagrangian}. {\bf Left:} In {\bf blue} we show the ratio of the cross section for the $e^+_{\text{vis}} e^-_{\text{vis}} a$ final state vs $\gamma_{\text{vis}} a$ final state as a function of the angular acceptance of a hypothetical lepton collider with center of mass energy $\sqrt{s}=10.58\text{ GeV}$. In {\bf yellow} we show the ratio between the two processes contributing to $e^+_{\text{vis}} e^-_{\text{vis}} a$: the ``ALP Dalitz'' process where a photon is exchanged in $s$-channel and the ``photon-fusion'' into ALP where the photon is exchanged in $t$-channel. The {\bf red dashed} line shows the \BelleII minimal opening angle for photons and electrons/positrons to be detected by the calorimeter. The {\bf red dotted} line shows the \BelleII minimal opening angle for electrons/positrons to be detected by the tracker. {\bf Right:} ALP cross sections at \BelleII as a function of the ALP mass $m_a$. The mono-photon ALP cross section ({\bf blue}) is roughly 10 times {larger} than the photon-fusion cross-section with final state $e^+$ and $e^-$ in the \BelleII acceptance one ({\bf yellow}). We also show the {\it inclusive} $e^+e^- a$ ({\bf {green}} line) which is orders of magnitude larger than $e^+_{\text{vis}}e^-_{\text{vis}} a$ due to the acceptance of the \BelleII detector. The {\bf green} band indicates the EPA approximation of $e^+e^- a$ with its associated uncertainty.}
    \label{monog_vs_eea}
\end{figure*}

\section{Signal}~\label{signal} 
Starting from the model in Eq.~\eqref{eq:ALPlagrangian} two possible ALP production mechanisms at a lepton collider are  
\begin{eqnarray}
    e^+e^-\to \gamma_{\text{vis}} a\,,\label{eq:alp-strahlung}\\  e^+e^-\to e^+_{\text{vis}} e^-_{\text{vis}} a\ ,\label{eq:signal}
\end{eqnarray}
where the subscript ``vis'' indicates that we require the photon and the electron-positron pair to be within the geometric acceptance of the detector. The first process is the well studied ALP-strahlung leading to the $\gamma+\text{invisible}$ signal. The second process leads to a $e^+e^-+\text{invisible}$ signal where two different topologies contribute to the total cross section: $i$)~the ``ALP-Dalitz'' process, given by the ALP-strahlung Eq.~\eqref{eq:alp-strahlung} with a photon conversion into $e^{+}e^{-}$\footnote{We remark that other conversions of the photon are possible, e.g. $\gamma^* \to \mu^+\mu^-$, and may provide further signal.}, $ii$) the ``photon-fusion'' into ALP, given by  a photon line exchanged in the $t$-channel that radiates the ALP. 

As shown in Fig.~\ref{monog_vs_eea} left, the angular acceptance controls the hierarchy between the $\gamma+\text{invisible}$ and the $e^+ e^- +\text{invisible}$ cross sections. Moreover, within the $e^+ e^- +\text{invisible}$ final state, the angular acceptance controls the hierarchy of the ALP-Dalitz vs the photon-fusion channels.

The angular acceptance of the \BelleII tracking system requires the polar angle of every electron in the center-of-mass frame, denoted by $\thetaestar$, to be more than $17\degree$ away from the beam axis. However, a successful electron ID should be supplemented by information from the ECAL. This raises the requirement on the minimal acceptance angle and introduces a further requirement on the minimal energy 
\begin{equation}
    \theta_{\min}^*=22\degree\ ,\qquad E^*_{\text{min}}=0.25\text{ GeV}\,,  \label{eq:missedgamma}
\end{equation}
in the center of mass frame. The same acceptance applies to photons, that are reconstructed mainly by the ECAL. The signal cross sections in the acceptance of  \BelleII with center-of-mass energy $\sqrt{s}=10.58\text{ GeV}$ are given in Fig.~\ref{monog_vs_eea}~right. They can be approximated for small enough ALP masses as 
\begin{equation}
    \begin{split}
        &\sigma(e^+e^-\to \gamma a)\approx10^{-3}\text{ pb}\left[\frac{g_{a\gamma\gamma}}{10^{-4}\text{ GeV}^{-1}}\right]^2\, ,\\
        &\sigma(e^+e^-\to e^+ e^- a)\approx7\times10^{-5}\text{ pb}\left[\frac{g_{a\gamma\gamma}}{10^{-4}\text{ GeV}^{-1}}\right]^2\!. 
    \end{split}\label{eq:signal-rate}
\end{equation}
As a result, the ALP-strahlung cross section is larger than the photon-fusion one by roughly a factor of 14 at \BelleII. As shown in  Fig.~\ref{monog_vs_eea}~left, the $e^+e^-\to e^+ e^- a$ cross section at \BelleII is still dominated by the photon-fusion channel,  although the Dalitz contribution is  only a factor of 5 smaller.

Both these facts are the result of the strong suppression of the inclusive $e^+e^-\to e^+ e^- a$ cross section due to the Belle~II acceptance. The photon-fusion rate is dominated by  electrons and positrons close to the beam axis, that unavoidably fall out of the \BelleII acceptance. Indeed, as it is shown in Fig.~\ref{monog_vs_eea}~left, extending the angular coverage down to $\theta_{\min}^*\simeq 1\degree$ the cross section of the $e^+e^-\to e^+e^- a$ process would become of the same size of the ALP-strahlung $e^+e^-\to a\gamma$, with the photon-fusion dominating over the Dalitz process by almost two orders of magnitude.

For comparison we show in Fig.~\ref{monog_vs_eea} right the inclusive cross section of  $e^+e^-\to e^+ e^- a$ which is more than two orders of magnitude larger than the one where the $e^+e^-$ pair is efficiently reconstructed at Belle~II. The inclusive cross section can be approximated using the effective photon approximation (EPA)~\cite{Fermi:1925fq,vonWeizsacker:1934nji,Williams:1934ad,Budnev:1975poe,Frixione:1993yw} which is shown as a green band in Fig.~\ref{monog_vs_eea} using the implementation of MadGraph~3.5~\cite{Alwall:2014hca} with the associated theoretical uncertainty corresponding to a variation of the factorization scale from $10\, m_a$ to $0.1\,m_a$. A more rigorous theory uncertainty could be associated following Ref.~\cite{Frixione:1993yw}. On the contrary, requiring the $e^+e^-$ pair to be reconstructed at \BelleII forces the kinematic of $e^+e^-\to e^+e^- a$ to depart substantially from the kinematics encompassed by the EPA approximation. A full analytical understanding of this $2\to3$ process in a general kinematic configuration will be presented in a forthcoming publication~\cite{future}.

\subsection{Signal kinematics}\label{sec:sigkin}

We now describe the signal kinematics for the two signal production modes of Eqs.~\eqref{eq:alp-strahlung} and \eqref{eq:signal}. In the mono-$\gamma$ topology the photon and the ALP momenta are back-to-back $\vec{p}_\gamma=-\vec{p}_{a}=-\vec{p}_{\text{miss}}$. Under these circumstances the signal is characterized  by a  photon of fixed energy $E_\gamma$ and the amount of missing energy $E_{\text{miss}}$ required by energy conservation. Explicitly one finds
\begin{equation}
    E^*_\gamma=\frac{s-m_a^2}{2\sqrt{s}}\,,\qquad E^*_{\text{miss}}=\frac{s+m_a^2}{2\sqrt{s}}\ , \label{eq:Emiss-monophoton}
\end{equation}
where $\sqrt{s}$ is the center of mass energy and the quantities defined in the center of mass frame (CoM) will be denoted with a $*$ throughout this work. 

After requiring a single photon of fixed energy up to the experimental resolution, a further discrimination between signal and background can only be achieved by selecting the central detector region. A standard way to do so is to define the missing pseudo-rapidity in the lab frame as
\begin{equation}
    \eta_{\text{miss}}\!=\!\frac{1}{2}\!\log\!\left[\frac{\vert\vec{p}_{\text{miss}}\vert+\pLmiss }{\vert\vec{p}_{\text{miss}}\vert- \pLmiss }\right]\!=\!-\log\left[\tan\frac{\theta_{\text{miss}}}{2}\right]\,\!\! ,\label{eq:missingeta}
\end{equation}
where we defined $\pLmiss$ as the component of the missing momentum along the beam pipe and $\theta_{\text{miss}}$ as the angle made by the ALP trajectory with the beam pipe axis. Selecting the central region is equivalent to requiring an upper bound $\vert\eta_{\text{miss}}\vert$. 

Due to its three-body nature, signal kinematics for the $e^+_{\text{vis}} e^-_{\text{vis}} a$ channel can be characterized by the missing mass $m_{\text{miss}}$ together with the $E_{\text{miss}}$ and $\eta_{\text{miss}}$. These quantities can be used to distinguish the ALP production from background processes.

The missing mass
\begin{equation}
    m_{\text{miss}}^2=E_\miss^2-|\vec{p}_\miss|^2=m_a^2\,,\label{eq:missing_mass}
\end{equation}
is equal to the ALP mass up to the experimental resolution. The missing energy and the missing momentum can be written as a function of the visible electron-positron pair in the final state and the CoM energy: 
\begin{equation}
    E_{\text{miss}}^*=\sqrt{s}-E_{e^+}^*-E_{e^-}^*\ ,\qquad \vec{p}_{\text{miss}}^*=-\vec{p}_{e^+}^*-\vec{p}_{e^-}^* . \label{eq:Emiss}
\end{equation}
Analogously to Eq.~\eqref{eq:missingeta} a missing pseudo-rapidity can be defined from the initial and final state electrons and positrons.\footnote{In principle further information can be extracted from the azimuthal angle between the positron and the electron in the final state.} 

\begin{figure*}
    \includegraphics[scale=0.35]{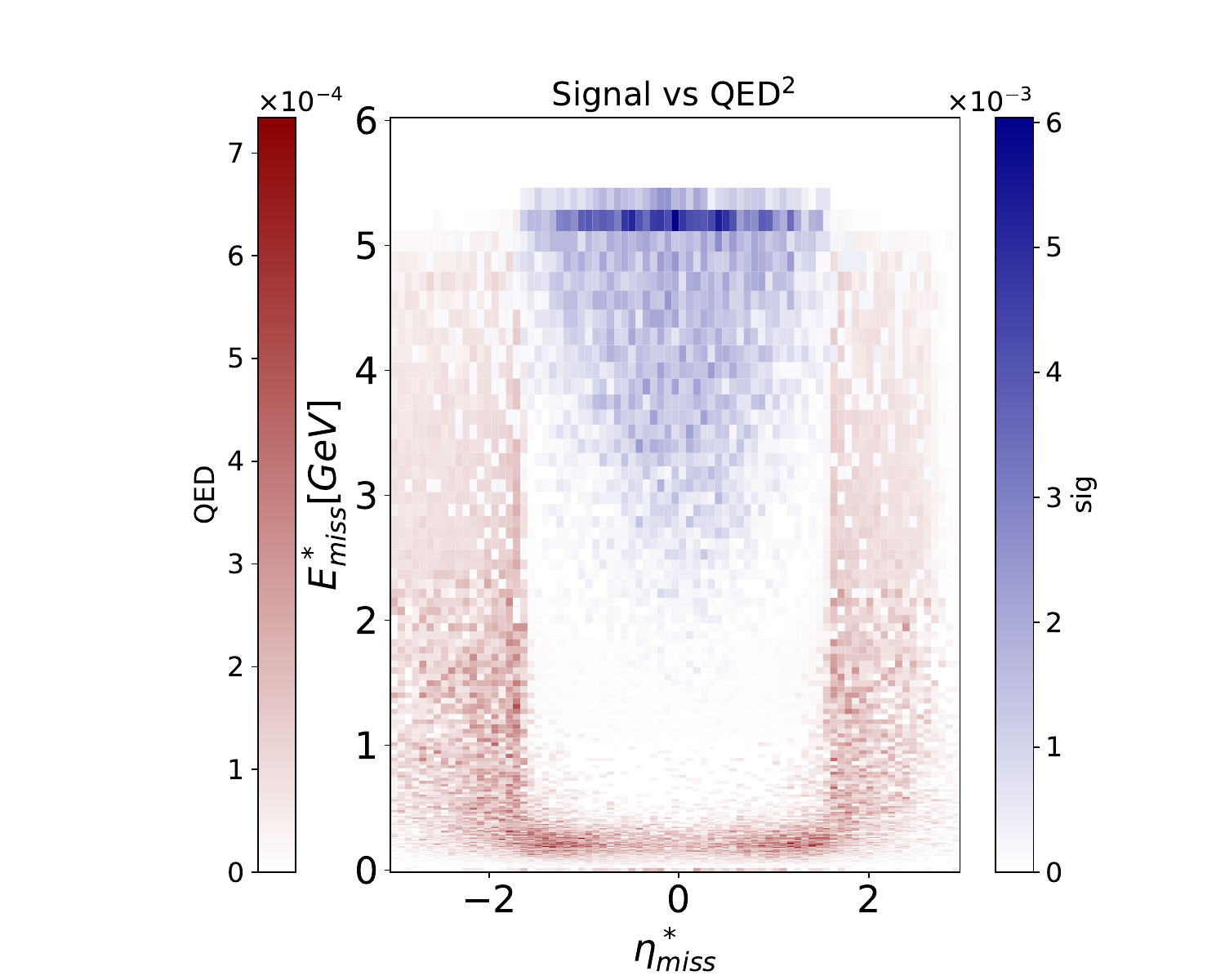}
    \includegraphics[scale=0.35]{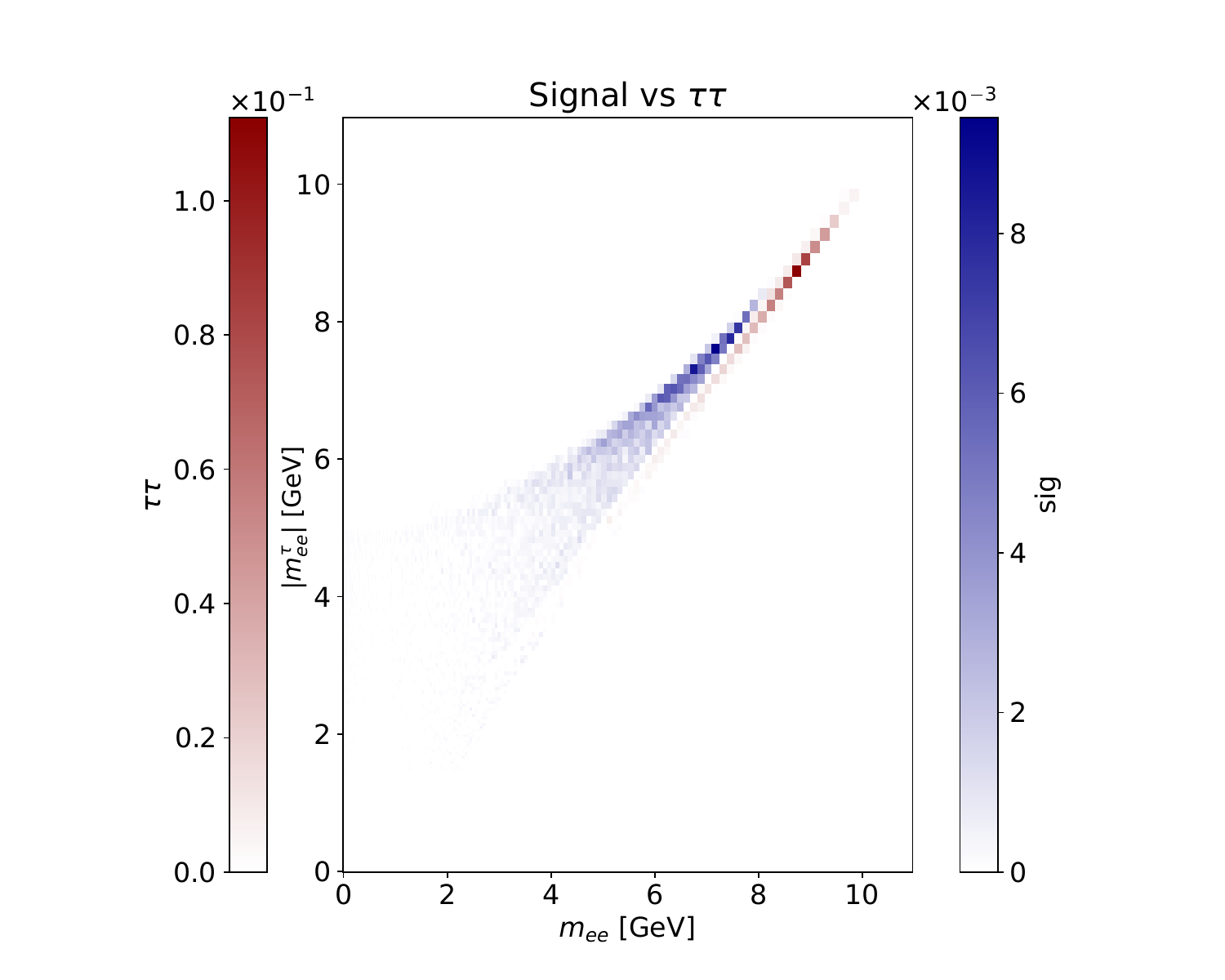}
    \caption{Distributions of the signal $m_a=1\text{ GeV}$ and the backgrounds ({\bf red}) with respect to the discriminating variables. In both panels we ask for the final electrons to be within Belle II acceptance and we select $| m_{\text{miss}}^2 - m_a^2|\leq \kappa \delta m_\miss^2$ with $\kappa$ chosen to maximize the sensitivity as described in Sec.~\ref{sec:event-selection}. In the {\bf left} panel we compare the signal to the QED$^2$ distribution with respect to $\eta_{\miss}^{*}$ and $E^*_{\miss}$ with uniform binning. In the {\bf right} panel we compare the signal to the $\tau\tau$ distribution with respect to $m_{ee}$ and $|m_{ee}^{\tau}|$ as defined in Sec.~\ref{sec:tautau}. The binning is such that $\delta E^*_{\miss}/E^*_{\miss}=2\%$, $\delta m_{ee}/m_{ee}=2\%$ and $\delta\eta^*_{\text{miss}}=0.075$ even though the experimental resolution at Belle~II can be better (see App.~\ref{subsec:en_angle_smearing}).}
    \label{empty_monog_ee}
\end{figure*}

Besides the fixed missing mass, the ALP signal is expected to be central (i.e. small $\eta_{\text{miss}}$) as confirmed by the distribution in Fig.~\ref{empty_monog_ee} left. Moreover, requiring the electron-positron pair within the Belle~II acceptance favors a kinematics where the ALP is not produced at rest resulting in a large $E_{\text{miss}}^*$, even for a very light ALP. For example for an effectively massless ALP we find that more than 90\% of the signal has $E_{\text{miss}}^*\gtrsim 2\text{ GeV}$ (see Fig.~\ref{fig:missvsacc}). Indeed, in Fig.~\ref{empty_monog_ee}~left  we can see the ALP signal at Belle~II is characterized by a large missing energy. This is at odds with the usual expectation for the fusion production mechanism, e.g. Higgs boson production at LHC in VBF~\cite{Rainwater:1999sd,Altarelli:1987ue}, because the phase space of the photon-fusion process producing the ALP at rest is cut out from the Belle~II geometric acceptance. In App.~\ref{app:signal} we comment further on how this feature depends on a hypothetical change of the angular acceptance of Belle~II.  

Leveraging all these distinctive features of the signal, we will argue that it is very difficult for any SM background to have a \emph{large} $E_{\text{miss}}$ together with a \emph{small} $\vert\eta_{\text{miss}}\vert$ and a \emph{small} $m_{\text{miss}}$. In the next section we will show how it is possible (especially for light ALP masses) to design a search at \BelleII where the background rejection is so good to compensate the suppressed production cross section with respect to the $\gamma+\text{invisible}$ channel illustrated in Eq.~\eqref{eq:signal}.

\section{Backgrounds}\label{qed-zero}

SM  processes  that give the same final state as our signal  Eq.~\eqref{eq:signal} are
\begin{align}
    & e^+ e^- \to e^+ e^-  \,+\, n \gammainv\, , \label{eq:qed_n_scat}\\
    & e^+ e^- \to \tau^- \tau^+, \tau^{\pm} \to e^{\pm}  \nu \nu\,,\label{eq:tau_scat}
\end{align}
where $\gammainv$ indicates a ``missed photon'' that cannot be detected at \BelleII because either its energy is below the ECAL energy threshold, or it is emitted too close to the beam direction and ends up in the two blind cones around the forward and backward direction. In formulas $\gammainv$ for \BelleII is defined as a photon that fails the requirement of Eq.~(\ref{eq:missedgamma}). This definition does not take into account further  detector inefficiencies which will be discussed in Sec.~\ref{sec:futher-backgrounds}. We will refer to Eq.~(\ref{eq:qed_n_scat}) as QED$^n$ background, for which the leading order QED in perturbation  theory corresponds to $n=2$. Eq.~(\ref{eq:tau_scat}) is referred in the following as the $\tau\tau$ background.

Both backgrounds above are \emph{reducible} thanks to the fact that their kinematics does not resemble at all the one of the signal.\footnote{An \emph{irreducible} SM background for the signal topology considered here would be the production of $Z$ bosons followed by their invisible decay into neutrinos, $Z\to \nu\nu$. Given the large mass of the $Z$ boson with respect to the \BelleII CoM energy this background is not going to be a concern for us.} However, both background processes have a huge cross section compared to the signal  Eq.~\eqref{eq:signal-rate}. In particular, given the Belle~II acceptance, the $\tau\tau$ background cross-section is around 75.4~pb and the QED$^{2}$ background is of 1.66~nb. These number set the challenge of the photon-fusion search which needs to achieve a background rejection at the level of $10^{-8}$ to be sensitive to $g_{a\gamma\gamma}\simeq 10^{-5}\GeV^{-1}$, where the target for resonant ALP-mediated freeze-out lies. 

Luckily, we will show in Sec.~\ref{sec:missedgamma} that it is possible to select a region of phase space that contains most of the signal where the QED$^n$ background is forbidden for kinematical reasons. In Sec.~\ref{sec:tautau} we discuss the $\tau\tau$-background and we take advantage of its peculiar kinematics to define different regions of phase space distinguished by their level of signal purity.

\subsection{QED$^n$ backgrounds}~\label{sec:missedgamma}

We explore here the kinematics of the QED$^n$ background where the missing energy is faked by an arbitrary number of $\gammainv$ not satisfying either the minimal energy threshold or the minimal angular acceptance defined in Eq.~\eqref{eq:missedgamma}. Given the energy and angular cuts at Belle~II these backgrounds can be studied in perturbation theory without the need of resumming large logarithms due to soft or collinear divergences. 

The goal is to quantify the difficulty for a kinematical configuration where the missing energy is given by $n$-missing particles which are either \emph{soft} or \emph{forward/backward} to fake a signal which has \emph{large} missing energy, \emph{small} missing mass and a small $|\eta^*_{\text{miss}}|$. In particular we will prove that for every $\eta^*_{\text{miss}}$ there is a \emph{minimal} missing mass $\bar{m}_{\text{miss}}^2(\eta_{\text{miss}}^*)$ that the QED$^n$ background can realize. 

In order to simplify our discussion we will set $\eta^*_{\text{miss}}=0$ from now on. This choice corresponds to the largest minimal missing mass that the QED$^n$ background can realize and will be indicated as $\bar{m}_{\text{miss}}^2$ for brevity. We also take all the photons azimuthal angles to be equal and we set them to zero for simplicity and without loss of generality~\cite{tesiA}.

Given the large amount of missing energy in the signal event, it is very difficult for soft $\gammainv$ to fake the signal. The only possibility would be to have $n$ soft photons such that $n E^*_{\text{min}}=E^*_{\text{miss}}$ and $\eta^*_{\text{miss}}=0$.
However, in practice the required $n$ is so large, of order $\mathcal{O}(E_{\miss}^*/E_{\min}^*)$, to make the rate of this process extremely suppressed. We can then ignore soft photons from now on. 

Similarly, in QED$^n$ background topology with only one hard $\gammainv$, the remaining $n-1$ soft photons should nearly cancel the $z$ component of the momentum of the hard photon, as to pull the resulting invisible body towards central rapidity. This process will be also highly suppressed in perturbation theory and can be neglected.  

The most dangerous configuration is the one with two hard photons undetected because they fly close to the beam axis with large longitudinal momenta in opposite directions. Such configurations feature {\it large} angular separation between the photons, hence $m_{\miss}$ is typically large. In particular, we can define the minimal invariant mass of the QED$^2$ background 
\begin{equation}
    \bar{m}_{\text{miss}}\vert_{\text{QED}^2}= E^*_{\text{miss}}\cos\theta^*_{\text{min}}\ ,   \label{eq:min2gamma}
\end{equation}
by sending one $\gammainv$ at an angle $\theta^*_{\text{min}}$ with respect to the beam and the other at $\pi-\theta^*_{\text{min}}$. This result tell us that the minimal invariant mass of a QED$^2$ background for $
\eta_{\text{miss}}=0$ and for $E_{\text{miss}}\gtrsim4\text{ GeV}$ is around $3.7$~GeV making it easy to separate ALP with $m_a<\bar{m}_{\text{miss}}\vert_{\text{QED}^2}$ from the QED$^2$ background. 

We now consider the QED$^3$ background. In particular we want to demonstrate that adding extra $\gammainv$ cannot substantially change the minimal invariant mass in Eq.~\eqref{eq:min2gamma} ensuring the robustness of our strategy. Fixing two hard photon along the ECAL edges ($\theta_1=\theta_{\text{min}}, \theta_2=\pi-\theta_{\text{min}}$) and adding a third photon of angle $\theta_3$ and energy $E_3$ the missing mass can be written as a deformation of Eq.~\eqref{eq:min2gamma}
\begin{widetext}
    \begin{equation}
        m_{\text{miss}}^2\vert_{\text{QED}^3}=\bar{m}_{\text{miss}}^2\vert_{\text{QED}^2}+2 E_3^* E_{\text{miss}}^*\sin\theta_{\text{min}}^*\Delta_3-E_3^{*2}\Delta_3^2\ ,   \label{eq:min3gamma}
    \end{equation}
\end{widetext}
where we defined $\Delta_3=\sin\theta_{\text{min}}^*-\sin\theta_3^*$. For soft photons $\Delta_3$ can take any sign and at leading order the invariant mass of the QED$^2$ process gets shifted by a correction which is however suppressed by $\mathcal{O}(E_{\text{min}}^*/E_{\text{miss}}^*)\lesssim 0.06$. Hard photons can escape the detection only for $\theta_3^*<\theta_{\text{min}}^*$ which implies that $\Delta_3>0$. In this regime $m_{\text{miss}}^2\vert_{\text{QED}^3}>\bar{m}_{\text{miss}}^2\vert_{\text{QED}^2}$. For this reason in our background simulation we focus on the QED$^2$ background, which should capture the kinematical features of the relevant QED backgrounds up to detector inefficiencies.

Fig.~\ref{empty_monog_ee} left clearly shows a separation of QED from ALP signal in the plane $(\eta^*_{\miss},E^{*}_{\miss})$. Both background and signal distribution are shown after the missing mass cut has been enforced. As can be seen by eye the QED background in red is hitting a kinematical boundary at  $E^{*}_{\miss}\simeq 1\GeV$ and $|\eta^*_{\miss}|$.

\subsection{$\tau\tau$ background}\label{sec:tautau}
For the $\tau\tau$ background we take advantage of the peculiar kinematics of the electron-positron pair which originates from the the two $\tau$ leptons being: $i$) on-shell resonances; $ii$) flying back-to-back. Indeed, one can understand $\tau\tau$ background as an antler topology~\cite{Han:2009ss,Franceschini:2022vck} and exploit results for this type of events. 

We describe each $\tau$ decay as the decay 
$$\tau^{\pm} \to e^{\pm} N_{\pm}\,,$$
where $N_{\pm}$  is a composite object made of the two neutrinos that appear in the decay of $\tau_{\pm}$. Being a composite object, the body $N_{\pm}$ has a mass given by $2 p_{\nu_{e}} \cdot p_{\nu_{\mu}}$, hence it changes event by event depending on the actual values of the four-momenta of the two neutrinos. We write the event-dependent mass of the body $N_{\pm}$ as $m_{N_{\pm}}$. With this notation in mind, the energies in the rest frame of the decaying $\tau^{-}$ are 
\begin{equation}
    \begin{split}
        & E^{\tau^-}_{e^-}=\dfrac{m_\tau^2+m_e^2-m_{N_{-}}^2}{2 m_\tau}\,,\\
        &\ E^{\tau^-}_{N_{-}}=\dfrac{m_\tau^2-m_e^2+m_{N_{-}}^2}{2 m_\tau}\,.
    \end{split}
    \label{eqs:tau_rest_energies}
\end{equation}
A similar set of equations can be written in the $\tau^{+}$ rest frame for the energies of $e^{+}$ and the composite body $N_{+}$. We now aim at finding  what phase-space will be {\it not} accessible to the invisible particles of the $\tau\tau$ process. The region of inaccessible phase-space will be used as a selection to remove the background from the $\tau\tau$ process.

We will approximate the energies in Eq.~\eqref{eqs:tau_rest_energies} in the limit of negligible $m_e$ and negligible $m_{N_{\pm}}$. Neglecting $m_{N}$ with respect to $\sqrt{s}$ may not be an accurate approximation. Nevertheless it useful for our purposes, because it enlarges the phase-space of the invisible bodies to their maximal size. Thus, in this approximation the region of inaccessible phase-space that we find is smaller than the one resulting from an exact computation. This makes our approximation conservative. 

Within this approximation, the kinematics of the $e^{\pm}$ and the $N_{\pm}$ can be written in the \BelleII CoM frame as function of just  3 quantities: $i$) the angles $\theta_{\pm}$ of the $e^\pm$ with respect to the direction of flight of the respective parent $\tau$ lepton; $ii$) the angle $\phi$ between the planes of the decay products of the $\tau^\pm$. Denoting with $\Lambda_{\pm}$ the Lorentz transformation that connects the $\tau^{\pm}$ CoM to the \BelleII  CoM we can write:
\begin{equation}\begin{split}
        & p_{e^-}=\dfrac{m_\tau}{2} \Lambda_- (1, s_-, 0, c_-)\ ,\\
        & p_{N_{-}}=\dfrac{m_\tau}{2} \Lambda_- (1, -s_-, 0, -c_-)\ ,\\
        & p_{e^+}=\dfrac{m_\tau}{2} \Lambda_+ (1, s_+ c_\phi, s_+ s_\phi, c_+)\ ,\\
        & p_{N_{+}}=\dfrac{m_\tau}{2} \Lambda_+ (1, -s_+ c_\phi, -s_+ s_\phi, -c_+)\, ,
    \end{split}\label{eqs:tau_momenta}
\end{equation}
where $p_\miss=p_{N_{-}}+p_{N_{+}}$ and we defined $s_x= \sin \theta_x, c_x=\cos \theta_x$ with $x \in \{+,-,\phi\}$ encompassing the polar angles of the $e^{\pm}$ and the angle between the decay planes introduced above.\footnote{We recall that the $\tau^{+}$ and the $\tau^{-}$ velocities are anti-aligned in the \BelleII CoM frame, therefore, once the $z$ axis is rotated along the direction of flight of the $\tau$ leptons the CoM of the $\tau^{-}$ is transformed in the CoM of \BelleII and in that of the $\tau^{+}$ by boosts along the $z$ direction of rapidity $y$ and $2y$, respectively, where $\cosh y=\sqrt{s}/2m_{\tau}$.} 

From the kinematic in Eq.~\eqref{eqs:tau_momenta} one can write down the invariant mass of the $e^{+}e^{-}$ pair originating from the $\tau\tau$ system and write the invariant mass as a function of $\theta_{+}$, $\theta_{-}$ and $\phi$. Defining the energies of electrons and positions in the \BelleII CoM as $E^*_{e^{\pm}}=\dfrac{\sqrt{s}\mp c_{\pm} \sqrt{s-4m_\tau^2}}{4}$ we can write 
\begin{widetext}
    \begin{equation}
        (m_{ee}^\tau)^2\!=\!\!\dfrac{2}{s-4 m_{\tau }^2}\!\!\left[m_{\tau }^4-\! \sqrt{s} m_{\tau}^2 (E_+^*+E_-^*)+\!2 E_-^* E_+^*(s-2 m_{\tau}^2)\!+\!m_{\tau }^2 c_\phi \mathcal{M}_{-} \mathcal{M}_{+}\right]\, , \label{eq:meetau}
    \end{equation}
\end{widetext}
where we defined $\mathcal{M}_{\pm}=\sqrt{m_{\tau }^2-2 E_\pm^* \sqrt{s}+4 E_\pm^{*2}}$. Since for the QED and the signal it can happen that $m_{\tau }^2-2 E_\pm^*\sqrt{s}+4 \left(E_\pm^*\right)^2<0$ we use the $|m_{ee}^{(\tau)}|$ in our selection described Sec.~\ref{sec:event-selection}.

As apparent from Fig.~\ref{empty_monog_ee}~right, the $\tau\tau$ background lives along a line of the space $m_{ee}-m_{ee}^\tau $. Up to possible mis-measurements of the electron and positron momenta the $\tau\tau$ background can be removed by filtering out events for which $m_{ee}=m_{ee}^\tau $. In Fig.~\ref{empty_monog_ee} we used a binning corresponding to $\delta m_{ee}/m_{ee}=2\%$. As the signal populates the upper part of the plane above the line $m_{ee}=m_{ee}^\tau $ it is possible to obtain high rejection of $\tau\tau$ while keeping a substantial amount of signal.

As discussed in Sec.~\ref{signal} our signal has both a $t$-channel dominated (photon-fusion) and an $s$-channel dominated (ALP-Dalitz) contribution. In Fig.~\ref{empty_monog_ee}~right we can identify a large-$m_{ee}$ region dominated by the $t$-channel photon-fusion and a small-$m_{ee}$ region dominated by the Dalitz process. 

The Dalitz dominated region at small $m_{ee}$ tends to be very well distinguishable from the $\tau\tau$ background, thus it may give rise to a background-free search, very safe from potential systematic uncertainties in the background estimation. Conversely, the fusion dominated region at large $m_{ee}$ lives close to the $\tau\tau$ background and can only be efficiently separated thanks to the   precision in the invariant mass measurements of \BelleII.

\subsection{Further backgrounds\label{sec:futher-backgrounds}}

We conclude this section discussing possible further backgrounds which are not included in the projected sensitivity in Fig.~\ref{fig:money} since their rate is difficult to estimate without a full comprehension of the detector performances. 

One background can arise from partial reconstruction of microscopic fully visible 2-body processes, e.g. 
$$e^+e^-\to q \bar{q}\to \text{hadrons}\to e^{+}e^{-}+\text{inv.}\,,$$
resulting in two isolated electrons and missing momentum. These kind of backgrounds arise because of the chance that a quark appears in the detector as an isolated charged lepton, e.g. because the leading hadron that appears from the color charge of the quark carries the vast majority of the quark momentum and decays semi-leptonically, that is $q 	\rightarrowtail \pi +\text{soft-particles} \to e\nu$. These backgrounds can be assimilated to the $\tau\tau$ background in that the unobserved particles of each of $q$ and $\bar{q}$ can be treated as a kind of invisible body for a suitable event-dependent mass analogous to $m_{N_{\pm}}$. Accounting for these backgrounds would roughly amount to a variation of the total rate of the $\tau\tau$ process after our selection.

In relation to the QED background the kinematical argument in Sec.~\ref{sec:missedgamma} is quite robust with respect to modification of the minimal angle $\theta_{\text{min}}^*$ or the minimal detectable energy $E_{\text{min}}^*$ at Belle~II. Vice-versa, losses of central photons around small non-instrumented regions of the detector are a potential source of very large backgrounds for our signal. For instance a lost $\gamma$  can give rise to a final state $e^{+}_{\text{vis}}e^{-}_{\text{vis}}\gamma_{\text{lost}}$, which is QED$^1$ in our power counting of the background processes. 

In such a case the measured $E_{\miss}$ could be due to this lost central photon recoiling against the detected $e^{+}e^{-}$ pair. Such configuration leads naturally to a small $m_{\miss}$, due to the small physical photon mass, and small $\eta_{\miss}^{*}$. In principle such background can be removed if $\eta^{*}_{\miss}$ is well measured by vetoing events for which the missing momentum falls in dead or not covered areas of the detector. A dedicated study of these background should be performed by the experimental collaboration. We further discuss this important caveat in Sec.~\ref{discussion}.

\section{Event selection and Sensitivity}\label{sec:event-selection}

We are now ready to summarize the kinematic selection we used to distinguish the signal kinematic described in Sec.~\ref{sec:sigkin} from the SM backgrounds described in Sec.~\ref{sec:missedgamma} and Sec.~\ref{sec:tautau}.  

A first discrimination between signal and backgrounds can be obtained from a selection on the  \emph{missing mass} defined in Eq.~\eqref{eq:missing_mass} which we want to fix around the ALP mass $m_a$ under examination up to the experimental resolution. The missing mass is obtained from a cancellation of two positive terms, $E_{\miss}$ and $|p_{\miss}|$, thus it is  expected that when $m_{\miss}$ tends to zero its experimental uncertainty gets large. After detector effects are included as detailed in Appendix~\ref{subsec:en_angle_smearing}, a good fit for the resolution on $m_{\miss}$ is 
\begin{equation}
    \delta m^{2}_{\text{miss}}\simeq \left[1  -  \left(\frac{m_{\text{miss}}}{10 \text{ GeV}}\right)^4\right] \text{  GeV}^2  \label{eq:delta_m2} \,. 
\end{equation}
In our selection we require  
\begin{equation}
    | m_{\text{miss}}^2 - m_a^2| \leq \kappa \cdot  \delta m^2_\miss \label{eq:miss_mass_selection}\ ,
\end{equation}
where the parameter $\kappa$ controls the width of the missing mass window which has been optimized to maximize the sensitivity.

We further characterize the signal kinematics demanding a large \emph{missing energy} $E_{\text{miss}}^{*}$ and a small \emph{missing rapidity} $\eta_{\text{miss}}^{*}$. In practice we require
\begin{align}
    &E^{*}_{\text{miss}} \in \left[E_{\text{miss}}^{\text{low}},\dfrac{s+m_a^2}{2 \sqrt{s}}\right]\ , 
    &|\eta_{\text{miss}}^*| \leq \eta_{\text{miss}}^{
        \text{high}}\,,
    \label{eq:emiss_selection}
\end{align}
where both $E_{\text{miss}}^{\text{low}}$ and $\eta_{\text{miss}}^{
    \text{high}}$ are chosen for each ALP mass in order to maximize the sensitivity. 

The cuts in Eq.~\eqref{eq:miss_mass_selection} and Eq.~\eqref{eq:emiss_selection} are chosen to optimize $S/\sqrt{B}$ in the cut-and-count scheme where $S, (B)$ indicates the number of signal (background) events. As an extra requirement we demand these cuts to keep at least $90\%$ of the signal. An example of the selection that we identify as optimal is given in Table~\ref{tab:event_selections} for two choices of ALP mass.

As shown in Fig.~\ref{empty_monog_ee} left the combination of Eq.~\eqref{eq:miss_mass_selection} and Eq.~\eqref{eq:emiss_selection} is enough to suppress most of the QED background. In addition, given the three-body nature of the signal, we can find a fourth selection variable to further improve the sensitivity. We find that the invariant mass of the visible $e^{+}e^{-}$ final state
\begin{equation}
    m_{ee}^{2}=2\,p_{e^{-}}\cdot p_{e^{+}}+2m_{e}^{2}=s+m_a^2-2\sqrt{s} E_a^*\,\label{eq:invmassee}
\end{equation}
is very effective to remove the background from $\tau\tau$ or any of its ``look-alike'' backgrounds described in Sec.~\ref{sec:futher-backgrounds}. The last equality in Eq.~\eqref{eq:invmassee} holds for the ALP signal only, whereas for the $\tau\tau$ background the ``antler'' topology implies that $m_{ee}$  should correspond to the value given in Eq.~(\ref{eq:meetau}).

This variable has also the merit of being a good discriminator between the Dalitz contribution to our signal, concentrated at low $m_{ee}$, and the fusion contribution concentrated at large $m_{ee}$. We then want to construct a test statistic which is able to incorporate both the \emph{low} $m_{ee}$ region which is in large part background-free and well-separated from the background and the \emph{high} $m_{ee}$ region where most of the signal cross section lies. In the latter case the separation between signal and background relies crucially on the resolution on $m_{ee}$ which distinguishes the $\tau\tau$-like background, aligned in the region where $m_{ee}\sim m_{ee}^\tau$, from the photon-fusion signal. 

\begin{table}[t!]
    \begin{tabular}{c|c|c|c}
        $m_a$ [GeV] & $\eta_{\text{miss}}^{\text{high}}$ & $E_{\text{miss}}^{\text{low}}$ [GeV] & $\kappa$\\ \hline 
        0.025 & 1.4 & 1.8 & 2.8 \\ 
        7 & 2.5 & 7.1 & 2 \\ 
    \end{tabular}
    \caption{Event selection parameters for two example ALP masses. $\eta_{\text{miss}}^{
            \text{high}}$ is the missing rapidity cut as in Eq.~(\ref{eq:emiss_selection}); $E_{\text{miss}}^{\text{low}}$ is the missing energy lower bound as in Eq.~(\ref{eq:emiss_selection}); $\kappa$ controls the width of the missing mass cut as in Eq.~(\ref{eq:miss_mass_selection}).}
    \label{tab:event_selections}
\end{table}

In practice, we construct a log-likelihood using the expected signal and background counts for 50~ab$^{-1}$ at \BelleII
\begin{equation}
    \Lambda=-2 \sum_{i,j} \ln \frac{L(S_{i,j},B_{i,j})}{L(0,B_{i,j})} \label{eq:loglik} \,,
\end{equation}
where $i$ and $j$ run on the bins of the plane  $(m_{ee},|m_{ee}^\tau|)$ as drawn in Fig.~\ref{empty_monog_ee}. For the computation of the likelihood we define bins of  width $\delta m/m=2\%$ motivated by the expected Belle~II resolution (see App.~\ref{app:bkd} for details). In   each bin we compute the Poisson factor
\begin{equation}
    L(S,B)=\frac{\left(S+B\right)^{B}}{B !} e^{-(S+B)}\,.
\end{equation}
The  sensitivity shown in Fig.~\ref{fig:money} corresponds to 95\% C.L. and it is obtained by requiring $\Lambda<4$. 

Some remarks on the robustness of our result are in order. The photon fusion channel contribution to $e^+e^-+\text{inv.}$  is potentially in danger of suffering from $\tau\tau$ background spillover in the signal-rich bins. This spill-over in real data may be due to resolutions effects. Our choice of $2\%$ resolution in the definition of the bins for the computation of the likelihood Eq.(\ref{eq:loglik}) should protect us from this type of problems. Furthermore in Fig.~\ref{fig:bkdextra} in the Appendix we show that our separation of the $\tau\tau$ background from the signal does not rely on a overly fine measurement of $m_{ee}$ and $m_{ee}^\tau$. To ensure the robustness of our sensitivity, we estimate the uncertainty due to finite MC sample performing several independent generations of our background MC. The variation of the  sensitivity over these replicas is negligible on the log-scale of our Fig.~\ref{fig:money}, and therefore not shown in the figure.

\section{Discussion}~\label{discussion}

In this work we derived the expected sensitivity on an ALP decaying invisibly at \BelleII with 50 $\text{ab}^{-1}$ of data in the channel $e^{+}e^{-}+\text{invisible}$. As shown in Fig.~\ref{fig:money} our study demonstrates that there is potential to improve significantly over the results based on the  $\gamma+\text{invisible}$ final state~\cite{BaBar:2008aby,BaBar:2010eww,BaBar:2017tiz,Dolan:2017osp,Belle-II:2018jsg} over the whole mass range. As shown in Fig.~\ref{fig:money}, the expected improvement on the reach can cover  most of the allowed parameter space for DM freeze-out through ALP-mediated annihilations in the resonant regime (see Ref.~\cite{Dolan:2017osp} and the discussion in Sec.~\ref{sec:setup} for details). 

For light ALP masses we argued that the signal kinematics is easily distinguishable from the SM background due to the interplay between low \emph{missing mass} and large central \emph{missing energy}. Remarkably, the excellent resolution of the Belle~II invariant mass measurements makes the $e^+e^- + \text{invisible}$ search competitive also for heavy ALP masses. For very heavy ALP masses (i.e. $m_a \gtrsim 8\text{ GeV}$) the $e^+e^- + \text{invisible}$ search can fill the gap where the $\gamma+\text{invisible}$ search encounters trigger issues related to the very large rate of single photon events at low photon energies.

In order to get a fair comparison of our proposal with the projected reach of the $\gamma+\text{invisible}$ channel an important issue is the possibility of undetected hard central photons. As discussed in Sec.~\ref{sec:futher-backgrounds}, this could be due to small non-instrumented regions of the detector or to other unspecified detector inefficiencies. This background is likely to be an important issue for our kinematic selection, but we cannot reliably include it in our simulation. On the contrary Ref.~\cite{Dolan:2017osp} claims to account for this type of background using not public official Belle~II background simulations prepared for the \BelleII physics book~\cite{Belle-II:2018jsg}. Ref.~\cite{Zanitalk} confirms that events of this type constitute at present the main challenge of the $\gamma+\text{invisible}$ channel. A similar background will affect our sensitivity in a way that should be estimated by the experimental collaboration. We hope that this work will trigger such a study.

Finally we mention two interesting future directions. First, the study of the fusion production mechanism for invisible ALP should be extended also for off-shell production (see for example Ref.~\cite{Essig:2013vha} for a similar study for $\gamma+\text{inv.}$ final state). In this case many of the kinematical cuts discussed here should be revised. Second, the importance of the production channel considered here should also carry over in many other experimental setups, including other high-intensity $e^+e^-$ colliders as well as future colliders. Significant differences can arise in the latter case, as new backgrounds due to electroweak bosons arise (see Ref.~\cite{Ruhdorfer:2023uea} for a first related study in this direction). We defer this investigation to a future work.

\section*{Acknowledgments}
We thank Enrico Graziani and Torben Ferber for discussions about \BelleII physics. We thank Ennio Salvioni and Alberto Mariotti for feedback on the draft.

\appendix

\section{Extra information on the event selection}\label{subsec:en_angle_smearing}
In this appendix we collect further information supporting the logic of our signal selection. We start by detailing how the sensitivity of the invariant mass to resolution effects was derived. In Sec.~\ref{app:signal} we characterize the signal kinematics, in Sec.~\ref{app:bkd} we discuss the separation between the QED and $\tau\tau$ backgrounds and the signal.

We model the resolution on the measurements on $e^{\pm}$ and $\gamma$ with gaussians centered at the particle-level value and standard deviations given in Ref.~\cite{Belle-II:2018jsg}:
\begin{eqnarray}
    \nonumber
    \frac{\delta E}{E}&=&\sqrt{
        \left[\frac{0.066\%}{E/\GeV}\right]^2 + \left[ \frac{0.81\%}{(E/\GeV)^{1/4}}\right]^2 + \left[ 1.34\% \right]^2\,,
    }\label{eq:ECAL-res}
    \\
    \delta\theta&=&10^{-3}\ .
\end{eqnarray}

Assuming initial state electron and positrons to be perfectly well measured we have computed the expected resolution on $m_{\miss}$ in Eq.~\eqref{eq:delta_m2} by repeated computations of $m_{\text{miss}}$ in Eq.~\eqref{eq:missing_mass} upon random variations of the energies and angles for the relevant kinematic configurations. These results yielded the fit given in Eq.~(\ref{eq:delta_m2}). Similar results have been obtained producing unweighted events with MadGraph and applying our selection described in Sec.~\ref{sec:event-selection} on events that have undergone a gaussian smearing described above. A similar procedure have been applied to $m_{ee}$ and $m_{ee}^\tau$ as shown in Fig.~\ref{fig:bkdextra} right.

\subsection{Signal}~\label{app:signal}
\begin{figure}[t]
    \begin{center}
        \includegraphics[width=1\columnwidth]{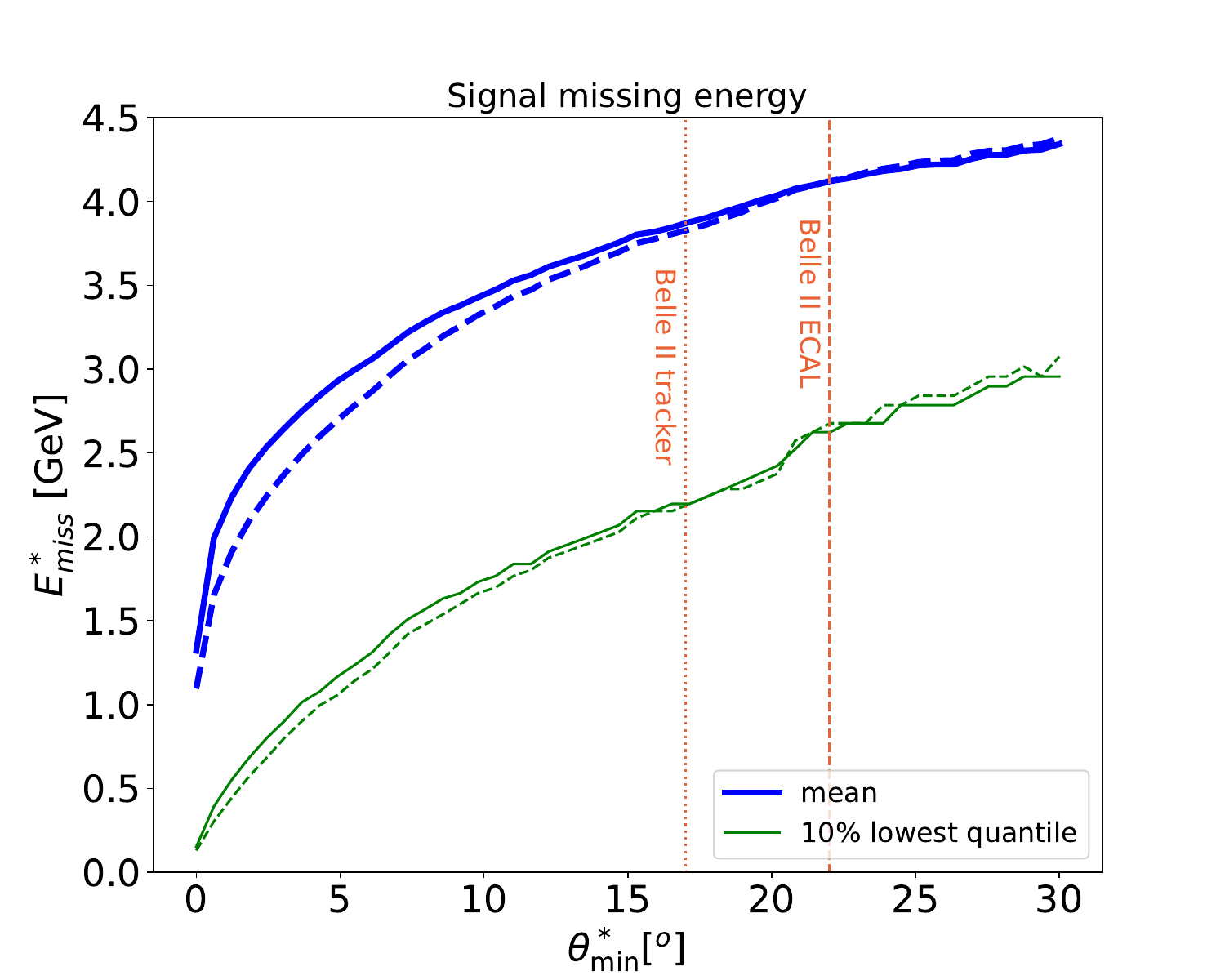}
        \caption{Missing energy for signal events of $e^+e^-\to e^+e^- a$ with $m_a=0.1\text{ GeV}$ as a function of the minimal angular acceptance of Belle~II. {\bf Thick blue } lines are sample averages. {\bf Thin green} lines are the 10\% lowest quantile of the sample. In the {\bf solid} lines no restriction on $\eta_\miss$ are imposed. In the {\bf dashed} lines $|\eta_\miss^*|<1.4$. The vertical {\bf red dashed} ({\bf red dotted}) line corresponds to the angular acceptance of Belle~II ECAL (tracking system) taken from Ref.~\cite{Belle-II:2018jsg}.}  \label{fig:missvsacc} 
    \end{center}
\end{figure}

Further features characterizing our signal can be understood from Fig.~\ref{fig:missvsacc} which shows how the average missing energy in the signal events depends on the minimal angular acceptance of Belle~II $\theta_{\min}^*$. We computed the average missing energy with no restrictions on the  missing rapidity (thick solid) or requiring $|\eta_{\miss}^*|<1.4$ (thick dashed). Comparing the two lines we find that this requirement does not significantly alter the  amount of $E_\miss$ in signal events. We also show as thin green lines the the 10\% lowest quantile of the sample. This gives the value of  $E_{\text{miss}}^*$ such that only $10\%$ of the signal sample has a smaller missing energy and can be used to understand the event selection of Table~\ref{tab:event_selections}. 

On a more speculative note, we notice that the angular acceptance of Belle~II was fixed in our work as to enforce the standard electron and photon reconstruction of Belle~II. This makes use of both ECAL and tracking information~\cite{Belle-II:2018jsg}. An interesting possibility would be to leverage the larger angular acceptance of the tracker going down to a $\theta_{\text{min}}=17\degree$ with a tracker-only identification of the positron and electron pair. Such a possibility does not exist for $\gamma+X$ signals, therefore it is an exploration specific of our signal and may lead to specific advantages for our final state. As shown in Fig.~\ref{monog_vs_eea}~left using only tracking for electron-ID will increase the signal cross section, thus potentially ameliorating the sensitivity. As discussed in Sec.~\ref{sec:futher-backgrounds} there are a number of possible backgrounds that can enter in our analysis if one considers less effective detection hardware, so the advantage from this loosening of the electron-ID must be carefully evaluated.

In general, changing the detector angular coverage may give large effects on the $E_\miss$. However, at a quantitative level Fig.~\ref{fig:missvsacc} shows that the average $E_\miss$ remains sizable even for rather small $\theta_{\min}^*$ around few degrees. Thus the preference for large $E_{\miss}$ of our signal, which we leveraged in our selection, keeps being an important and appreciable consequence of detector with significantly larger acceptance than the current Belle~II. From this consideration we conclude that imagining to instrument detectors in the forward region of Belle~II down to $\theta_{\min}=1-2\degree$ would increase the photon fusion cross section while maintaining the distinctive large missing energy in the signal events.

\subsection{SM backgrounds}\label{app:bkd}
\begin{figure*}[htbp]
    \begin{center}
        \includegraphics[scale=0.35]{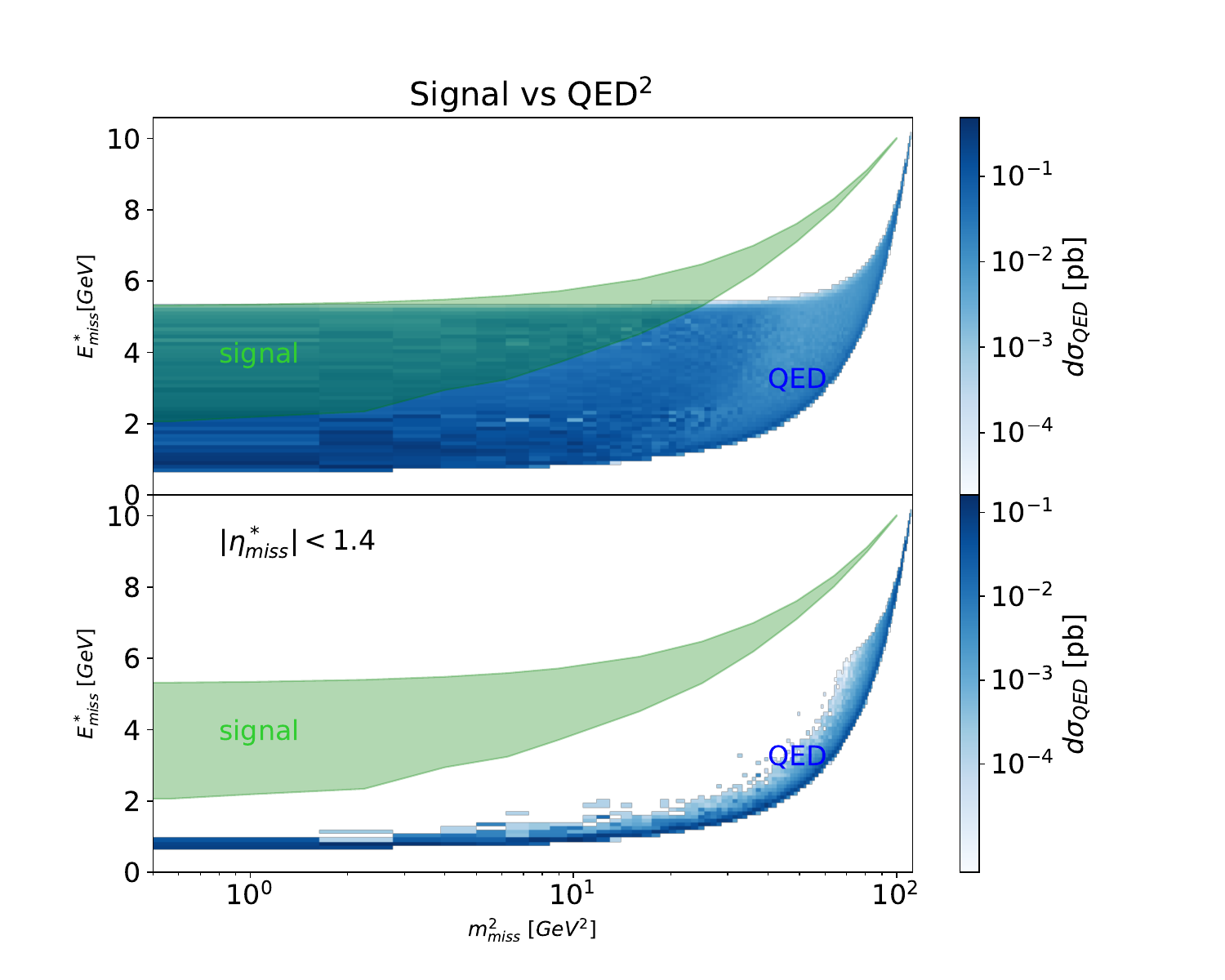}
        \includegraphics[scale=0.35]{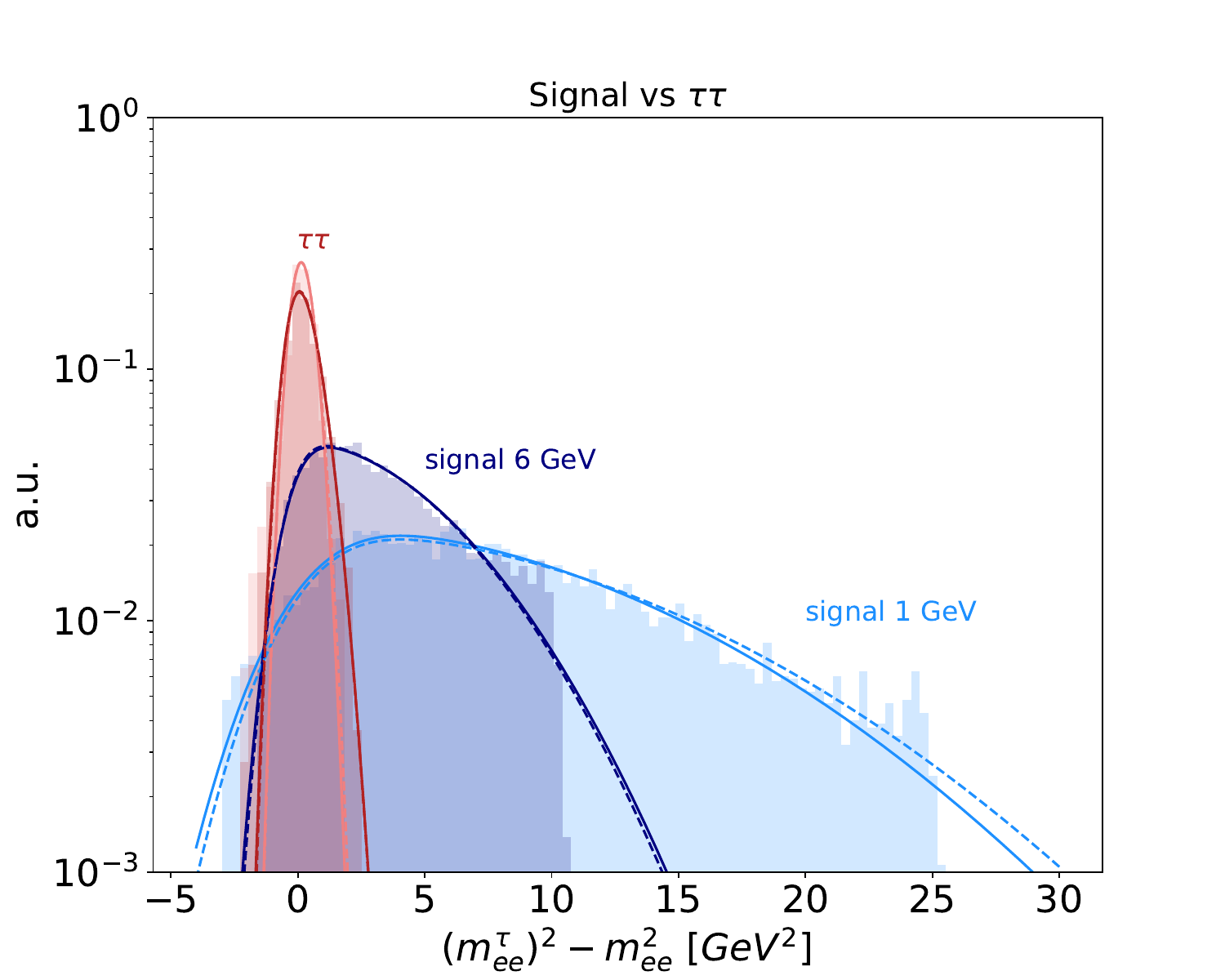}
        \caption{{\bf Left:} In {\bf blue} QED background distribution in the $(m_\miss^2,E_\miss^*)$ plane together with the $E_{\miss}$ selection Eq.~(\ref{eq:emiss_selection})
            for different ALP masses in {\bf green}. 
            In the top panel no restriction is applied to $|\eta^*_{\miss}|$, in the bottom panel we apply a selection 
            $|\eta_\miss^*|<1.4$, that is illustrative of the typical optimal requirement for central $|\eta_{\miss}^*|$.  {\bf Right:} Background distributions in {\bf red} vs signal 
            distribution in {\bf blue}. The separation between signal and background in $m_{ee}^{\tau}-m_{ee}$ is sharper for $m_{\text{miss}}=1\text{ GeV}$ ({\bf light} color palette) than for $m_{\text{miss}}=6\text{ GeV}$ ({\bf dark} color palette) where $m_{ee}^{\tau}$ is defined in Eq.~\eqref{eq:meetau}. Histograms are  MC data smeared according to Eq.~(\ref{eq:ECAL-res}). {\bf Solid lines} are fits with a skewed gaussian\footnote{The skewed gaussian non normalized pdf is $f(A,\mu,\sigma,\gamma,x)=A \exp \left(-\frac{(x-\mu)^2}{2 \sigma^2}\right) \left[1+\text{erf}\left(\frac{\gamma x}{2}\right)\right]$, with $A$ controlling the amplitude, $\mu$ the mean, $\sigma$ the standard deviation, and $\gamma$ the skewness parameter.} to smeared data; {\bf dashed lines} are fits to non smeared data. The comparison between the two fits shows the marginal impact of the smearing.}\label{fig:bkdextra}
    \end{center}
\end{figure*}

In Fig.~\ref{fig:bkdextra} we corroborate our analysis of the background kinematic in Sec.~\ref{qed-zero} that lead to our selection in Sec.~\ref{sec:event-selection}.

In the left panel we show the importance of an upper cut on $\eta_{\text{miss}}$ to select a signal kinematics which is not populated by the QED$^2$ background as defined in Sec.~\ref{sec:missedgamma}. Despite we focused our discussion on light ALPs Fig.~\ref{fig:bkdextra} left shows that the separation between the signal and QED$^2$ background is excellent for \emph{all} the ALP masses. This fact explain the flatness of our expected sensitivity in Fig.~\ref{fig:money} where the degradation at higher ALP masses can be ascribed to the reduced separation of the signal from the $\tau\tau$ background which we now explain.  

In the right panel we show how much our separation between the $\tau\tau$ background and the signal depends on the ALP mass. The signal normalized distribution is well spread in the $(m_{ee}, m_{ee}^\tau)$ plane with the Dalitz contribution concentrated at  small $m_{ee}$ and the fusion contribution at large $m_{ee}$. Increasing the ALP mass reduces the maximal $m_{ee}$ making the signal distribution closer to the background. This effect explain the degradation of our expected sensitivity at high ALP masses.

Contrary to the QED$^2$ background, the separation between the $\tau\tau$ background and the signal benefits from a good resolution in invariant mass. This is again clear from Fig.~\ref{fig:bkdextra}~left where the separation between the QED$^2$ and the signal can be seen ``by eye'' while the one between the $\tau\tau$ and the signal (right plot) very much depends on how precisely we can resolve the diagonal $m_{ee}\simeq m_{ee}^\tau$.

\bibliographystyle{JHEP}
\bibliography{bib}

\end{document}